\documentclass[aps,prl,twocolumn,showpacs,superscriptaddress,floatfix,nofootinbib]{revtex4}
\usepackage{graphicx} 
\usepackage[english]{babel}
\usepackage{amssymb}
\usepackage{amsmath}
\begin{document}
\newcommand{\p}{\prime}
\newcommand{{\ud}}{{\phantom1}}  
\newcommand{\bh}{\mbox{\scalebox{.9}{$\bullet$}}}  
\newcommand{\bb}{\mbox{\scalebox{0.5}{$\blacksquare$}}}  
\newcommand{\bd}{\mbox{\rotatebox{45}{\scalebox{0.5}{$\blacksquare$}}}}  

\title{Experimental probes of emergent symmetries in the quantum Hall system} 

\author{C.A. L\"utken}
\affiliation{Theory Group, Department of Physics, University of Oslo}
\affiliation{CERN, CH-1211 Geneva 23}
\author{ G.G. Ross}
\affiliation{CERN, CH-1211 Geneva 23}
\affiliation{Rudolf Peierls Centre for Theoretical Physics, Department of Physics, University of Oxford}
\date{\today}
\preprint{CERN-PH-TH-2010-192}
\preprint{OUTP-10-23P}
\begin{abstract}
Experiments studying renormalization group flows in the quantum Hall system provide significant evidence for the existence of an emergent holomorphic modular symmetry $\Gamma_0(2)$.  We briefly review this evidence and show that, for the lowest temperatures, the experimental determination of the position of the quantum critical points agrees to the parts \emph{per mille} level with the prediction from $\Gamma_0(2)$. We present evidence that experiments giving results that deviate substantially from the symmetry predictions are not cold enough 
to be in the quantum critical domain. 
We show how the modular symmetry extended by a non-holomorphic particle-hole duality leads to an extensive web of dualities related to those in plateau-insulator transitions, and we derive a formula relating dual pairs $(B,B_d)$ of magnetic field strengths across any transition. 
The experimental data obtained for the transition studied so far is in excellent agreement with the duality relations following from this emergent symmetry, and  rule out the duality rule derived from the  ``law of corresponding states". 
Comparing these generalized duality predictions with future experiments on other transitions should provide 
stringent tests of modular duality deep in the non-linear domain far from the quantum critical points.
\vskip .5cm\noindent CERN-PH-TH-2010-192/OUTP-10-23P
\end{abstract}
\maketitle

\section{I. INTRODUCTION}

The possibility that the quantum Hall (QH) system should possess a discrete non-Abelian emergent symmetry 
relating the complex conductivity $\sigma = \sigma_H^{\phantom D} + i\sigma_D^{\phantom D}$ 
in different QH phases was suggested by us some time ago\,\cite{Oxford1}, in order to explain the apparent ``super-universality" found in transitions between both integer and fractional levels in the QH system. The structure of the group emerged from the need to incorporate dualities between dyonic charges, as well as the periodicity associated with the topology of an effective theory encoding the anyonic nature of the quasi-particles. 
Together this led to the holomorphic modular symmetry $\Gamma_0(2)$. The symmetry was completed to a group that we shall call
 $\Gamma_{\rm H}$, through the inclusion of a particle-hole duality that we study in detail below.\footnote{The technical definition of the quantum Hall symmetry $\Gamma_{\rm H}$ involves the group of modular automorphisms, discussed in the next section.}
 
$\Gamma_{\rm H}$ successfully predicts the full phase diagram of the QH system, both integer and fractional, including the position of the quantum critical points governing the scaling behaviour of transitions between QH levels, as well as the scaling exponents.
A physical interpretation of this symmetry, based on the interchange of quasi-particles (describing the plateaux) and vortices 
(describing the QH insulator), was developed in ref.\,\cite{Cliff}.

In a related approach\,\cite{KLZ} based on an effective field theory incorporating  ``charge-flux transformations", a set of rules relating  
QH states at different filling factors, known collectively as ``the law of corresponding states", was constructed.  
This also determines the topology of  the phase diagram, but neither the location of quantum critical points, nor the geometry 
of renormalization group (RG) flows in the complex conductivity plane was obtained.   It also contains a kind of duality\;\cite{Tsui1, nu_duality}, 
but one that differs substantially from the modular duality contained in the symmetry $\Gamma_{\rm H}$.

We first discuss some of the substantial experimental evidence for the modular symmetry $\Gamma_{\rm H}$. 
Not only does the symmetry describe the properties of the observed integer and fractional QH plateaux, 
it also determines the temperature driven RG flow of the system\,\cite{Oxford2}. 
As briefly reviewed below, the position of the unstable fixed points and the RG trajectories are in good 
agreement with many experiments, particularly those involving the lowest temperatures, and so are the critical exponents. Indeed the one experiment that has been conducted at extremely low temperatures, an order of
magnitude lower than previously available\,\cite{Tsui_09050885},  confirms the fixed point structure predicted by the symmetry to very high accuracy.  We also discuss the experimental cases that are in apparent disagreement with the $\Gamma_{\rm H}$ 
symmetry and provide evidence that this is because the experiments are not cold enough,  i.e., that they are not probing the scaling domain.  

In addition to the temperature driven flows there is a significant body of experimental work studying the magnetic field driven transitions between the 
QH insulator (QHI) and both integer (IQH) and fractional (FQH) plateaux.   In most cases this flow is 
consistent with the flow predicted by the holomorphic emergent symmetry; the cases in apparent disagreement  are again conducted at relatively high temperature and it is very likely they are not probing the scaling domain. 

The plateau-insulator experiments also provide a sensitive test of the nature of the duality present in the quantum Hall system. The first direct experimental evidence for a duality symmetry in the QH system was
obtained over a decade ago\,\cite{Tsui1}, but there does not appear to have been
any experimental follow-up of this important discovery. The data were
immediately interpreted as evidence in favour of a new  ``charge-flux duality"\,\cite{Tsui1,nu_duality}.  
In subsequent sections we revisit the interpretation of this pioneering experiment,
and clarify the distinction between our approach and the charge-flux duality discussed in refs.\,\cite{Tsui1, nu_duality}. 
We are forced to conclude that it is the duality associated with the symmetry $\Gamma_{H}$ that is supported by the experiment,
and that the duality obtained from the law of corresponding states is, in fact, excluded.

Finally, in response to the advent of scaling experiments at extremely low temperatures, we work out
the duality relations for general transitions, particularly for the plateau-insulator transitions.  We also discuss the associated prediction of the dual value of the magnetic field corresponding to the dual QH state. Experiments measuring the dual pair properties should
provide stringent tests of the emergent symmetry $\Gamma_{\rm H}$, and in particular modular duality.

\section{II.  MODULAR SYMMETRIES}

We start with a brief review of the modular group $\Gamma_{\rm M} = {\rm PSL}(2,\mathbb{Z})$ and the subgroup 
$\Gamma_{0}(2)\subset \Gamma_{\rm M}$.  The modular group acts on the complex conductivity 
$\sigma = \sigma_H^{\phantom D} + i\sigma_{D}^{\phantom D} = \sigma_{xy}+i\sigma_{xx}$ by fractional
linear transformations generated by translations $T(\sigma)=\sigma+1$,
and by the ``complexified Kramers-Wannier duality"  transformation, $S(\sigma)= -1/\sigma$. 
Another representation of modular transformations is provided 
by integer valued matrices with unit determinant:
\begin{equation} 
\gamma = \begin{pmatrix} a & b\\c & d\end{pmatrix}\;\;,\;\;  \det \gamma = ad-bc  = 1\;,
\label{eq:modulartrafo}
\end{equation}
that act on the complex upper half-plane by the M\"obius transformations:
\begin{equation*} 
\gamma(\sigma) = \frac{a\sigma + b}{c\sigma + d}\;\;,\;\;
\sigma\in\mathbb{H} = \{\sigma\in\mathbb{C}|\Im\sigma = \sigma_D^{\phantom D} > 0\}.
\end{equation*}
Any transformation contained in the modular group $ \Gamma_{\rm M} = \{T,S\}$ is called modular 
and is a string or ``word" written with the letters $T$ and $S$.  Since $T$ and $S$ do not commute 
$\Gamma_{\rm M}$ is infinite, discrete and non-abelian.
There are two (and only two) independent relations\footnote{These ``grammatical rules" derive from 
the formal definition of the modular 
group as the free product of $\mathbb{Z}_2(S)$ and $\mathbb{Z}_3(TS)$:
$\Gamma_{\rm M}= \mathbb{Z}_2 \star\mathbb{Z}_3$.} 
satisfied by the generators, $S^2 = 1$ and $(TS)^3 = 1$, 
which can be used to simplify the group elements.   
$S$ has a fixed point of order two at $i = \exp(\pi i/2)$, and $TS$ has a fixed point of order three at $j = \exp(\pi i/3)$.  
Together with their images under the group these fixed points (to be identified with the quantum critical points 
associated with delocalization phase transitions) determine the topology and geometry 
of the phase and RG flow diagrams respecting modular symmetries\,\cite{Oxford1}.

\begin{figure*}[t]
\begin{center}
\includegraphics[scale = 1.1]{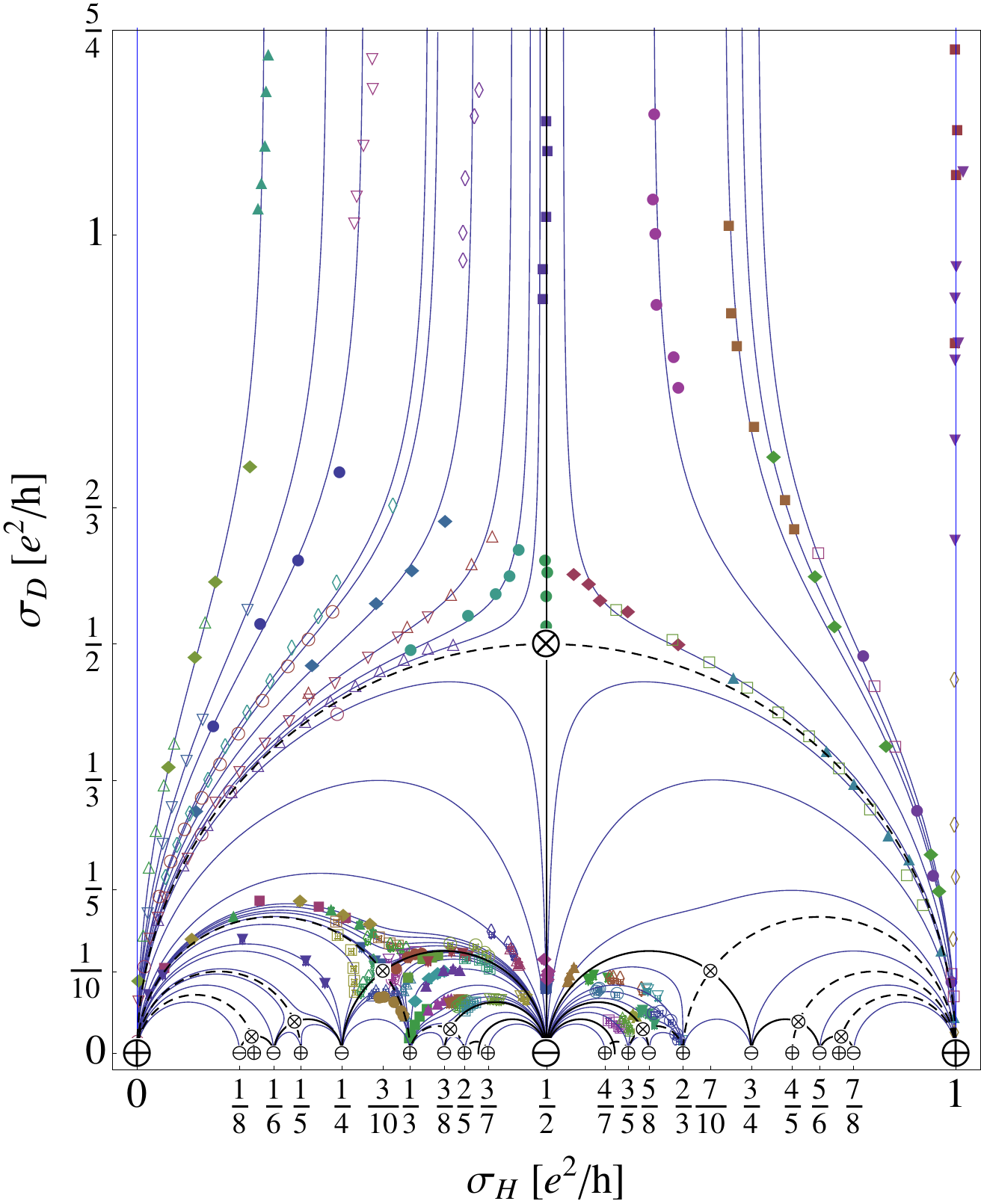}
\end{center}
\caption{(Color online) Conductivity phase- and RG flow diagram. Compilation of temperature-driven flow 
data\,\cite{Grenoble1,Grenoble2} superimposed on RG flow-lines derived from our 
$\Gamma_{\rm H}$-invariant RG potential. Thick black lines are phase boundaries.
Dashed lines are separatrices for the flow.}
\label{fig:Probes2-Fig1}
\end{figure*}

\begin{figure*}[t]
\par
\begin{center}
\includegraphics[scale = .6]{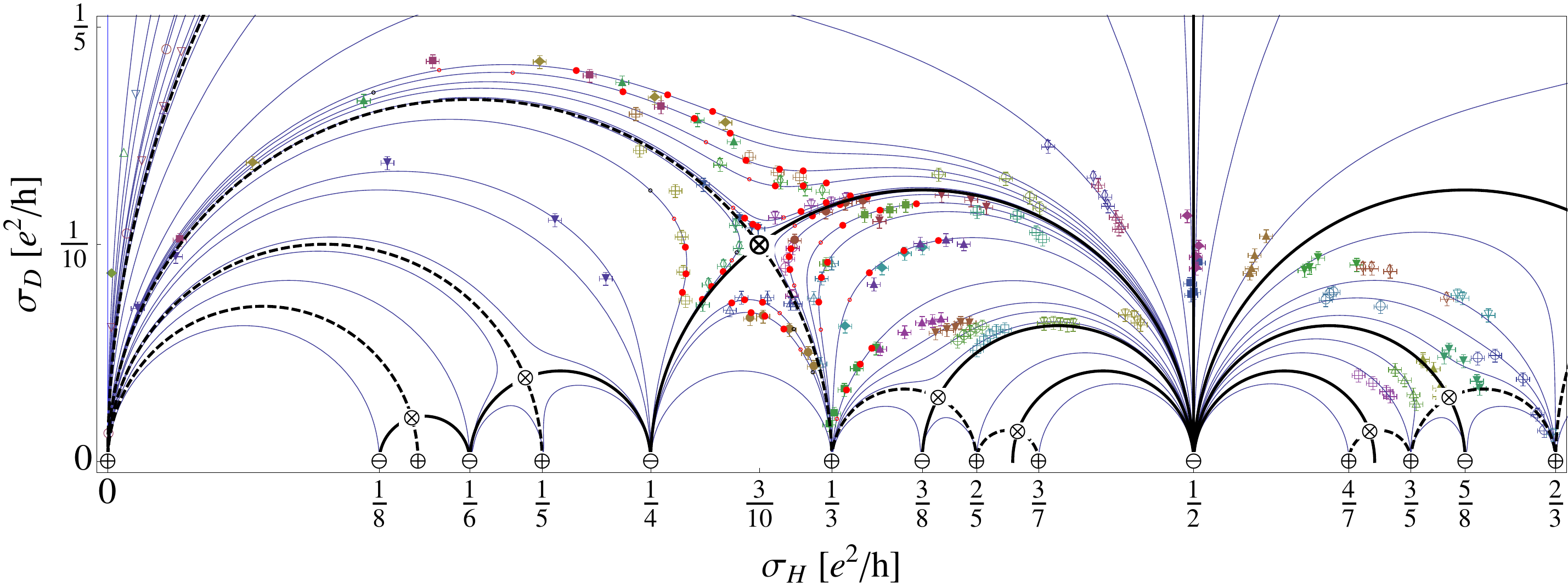}
\end{center}
\caption{(Color online) Comparison of experimental temperature-driven flows\,\cite{Grenoble2} 
with our theoretical model (thin blue lines). 
Real experimental data are equipped with error bars (discussed in ref.\,\cite{Oxford2}),
while all other icons are theoretical.  Round icons (red bullets and circles) are computed using the scaling exponents 
derived from $\Gamma_{\rm H}$.  For each flow line the first theory point is our choice of initial values for the flow, 
the rest are predictions.}
\label{fig:Probes2-Fig2}
\end{figure*}

The holomorphic part of the emergent symmetry group, when acting on the conductivities of the spin-polarized QH system, 
appears to be the discrete sub-group $\Gamma_{0}(2) = \{T,D\}\subset  \Gamma_{\rm M}$, 
which is generated by the translation $T$ and the  duality transformation $D(\sigma)=ST^{2}S(\sigma)=\sigma/(1-2\sigma)$.

The Hall bar is, effectively, a two-dimensional hetero-structure of size
$L_x \times L_y$, aligned with a current $I$ in the $x$-direction so that $R_{*x} = (L_*/L_y) \rho_{*x}$.  
The Hall resistance $R_H = \rho_H^{\phantom D} = \rho_{yx} = - \rho_{xy}$ is quantized in the fundamental unit  of resistance 
$h/e^2 \approx 25.81 \;k\Omega$,  while the dissipative resistance $R_D = \rho_{xx}/\square$  is rescaled by the 
aspect ratio $\square = L_y/L_x$.  The modular transformation $S$ conveniently relates complex conductivity 
and complex resistivity: 
\begin{equation*}
\rho= \rho_{xy} + i \rho_{xx} =  -\rho_H^{\phantom D} + i \rho_D^{\phantom D}  = S(\sigma) = - \frac{1}{\sigma}\;\;.
\end{equation*}
Hence the group $\Gamma^0(2)$ of symmetries acting on resistivities is $S$-conjugate to $\Gamma_0(2)$:
\begin{equation*}
\Gamma^0(2) = S\Gamma_0(2)S^{-1} = \{STS^{-1}, SDS^{-1} = T^2\} \;\;.
\end{equation*}
Since $S$ is not a symmetry of the spin-polarized QH system the conductivity group
$\Gamma_0(2)$ is not the same as the resistivity group $\Gamma^0(2)$.
In both cases the group acts holomorphically, which means that the symmetry transformations 
do not mix holomorphic coordinates $\sigma$ (or $\rho$) with anti-holomorphic (complex conjugate) coordinates
$\bar{\sigma}$ (or $\bar{\rho}$) - \emph{i.e.}, they respect the \emph{complex structure} of the model.

As originally proposed in ref.\,\cite{Oxford1} the system is invariant under an additional
symmetry  $J$ that flips the sign of $\rho_H^\ud$ ($\rho_D^{\phantom D}$ must be positive):
\begin{equation*}
J(\rho)=\sigma_{3}^{\phantom 3}(\bar{\rho})=-\bar{\rho}\;\;,
\label{eq:J}
\end{equation*}
where $\sigma_3^{\phantom 3}$ is the third Pauli
matrix. Note that this fractional linear transformation is not holomorphic,
nor is it modular since $\det\sigma_{3}^{\phantom 3} = -1$. Geometrically, $J$ is a reflection in the vertical axis, 
and it is the only generator of automorphisms of the upper half plane that is not modular. 
Physically, $J$ is a kind of ``particle-hole duality".
When this ``outer automorphism" is included the proposed symmetry group
for the fully spin-polarized QH system is\,\cite{Oxford1}:
\begin{equation*}
\Gamma_{\rm H} = \mathrm{Aut}\,\Gamma_0(2) = \{T,D,J\} \subset {\rm Aut}\, \Gamma_{\rm M} = \{T,S,J\}
\end{equation*}
when acting on conductivites, and $\Gamma^{\rm H} = \mathrm{Aut}\,\Gamma^0(2)$ when acting on
resistivities.  This is the \emph{emergent symmetry} that we shall demonstrate below is in detailed 
agreement with the experimental measurements.

\section{III.  EXPERIMENTAL TESTS OF $\Gamma_{\rm H}$}

There is by now considerable experimental evidence in favor of  $\Gamma_{\rm H}$ as a symmetry 
of the spin-polarized QH system\,\cite{Oxford2}. Much of this has been discussed before so here we present a brief summary.
Part of the conductivity phase and RG flow diagram is shown in Figs.\,\ref{fig:Probes2-Fig1} 
and \ref{fig:Probes2-Fig2}\,\cite{Oxford2}.   As originally noted in ref.\,\cite{Oxford1} the location of the fixed points
associated with the integer and fractional QH states are consistent with a $\Gamma_{\rm H}$-symmetric RG flow. 
This does not only include the attractive infra-red fixed points (we use the icon $\oplus$ to denote these points) 
associated with the plateaux, which appear automatically and inevitably with this symmetry, 
but also the location of the unstable quantum critical points (the delocalization fixed points, labelled by the icon $\otimes$) 
far removed from the attractors (in fact infinitely far away from the plateaux in the only natural metric, which is hyperbolic).
The measured geometry of the flow lines (data points, distinguished by having error bars) also appears to be in good agreement with the symmetry prediction (continuous lines), as is the rate of flow and the corresponding critical exponents.

Mapping out the phase diagram provides a crucial test of our model because the holomorphic modular symmetry is an  extremely rigid structure that cannot be modified in order to accommodate data. For this reason it is important to look at the accumulated evidence critically.
While the agreement of theory with experiment shown in Figs.\,\ref{fig:Probes2-Fig1} and \ref{fig:Probes2-Fig2} is impressive, 
there are some experiments that seem to disagree substantially with the structure predicted by $\Gamma_{\rm H}$.  In the next subsection
we address apparent discrepancies with the location of the fixed points and demonstrate that the temperature driven flow has not yet accessed the fixed point. In the second subsection we discuss some new data that {\it is} able to determine the fixed point location with high precision and show that its value agrees very closely with the predicted value.
The third subsection is devoted to a discussion of some experimental 
data that is in apparent disagreement with the prediction for the RG flow. A closer look at the data again indicates that this is because the quantum regime has not yet been reached.

The recurrent theme of this section is that the hyper-sensitivity to temperature exhibited by magneto-transport response functions
appears to have received insufficient attention in many cases.  In fact, we find that all differences between experiments and 
$\Gamma_{\rm H}$ disappear at sufficiently low temperatures, and that the data obtained at the lowest temperatures (tens of $mK$)
provide compelling evidence in favor of  $\Gamma_{\rm H}$.

\subsection{A.  Plateau-insulator quantum critical points}

The data shown in Figs.\,\ref{fig:Probes2-Fig1} and \ref{fig:Probes2-Fig2}  are in good agreement with the predicted location of the delocalisation fixed points, denoted by the icon $\otimes$ in our diagrams.  
However, there are some experiments on the QHI-IQH transition\,\cite{Tsui_9810217, Tsui_9906212} 
that appear to find critical magnetic fields and associated critical conductivities that are significantly 
displaced from the positions predicted by $\Gamma_{\rm H}$.  The critical magnetic field is 
identified\,\cite{Tsui_9810217,Tsui_9906212} 
as the stationary value of $B$ with respect to changing temperature.
The critical fields proposed by these investigators are far from the peaks of the curves $\sigma_D^{\phantom D}(B)$, 
which is where $\Gamma_{\rm H}$ predicts that the fixed point should be in this case.  

\begin{figure}[t]
\begin{center}
{\includegraphics[scale = .55]{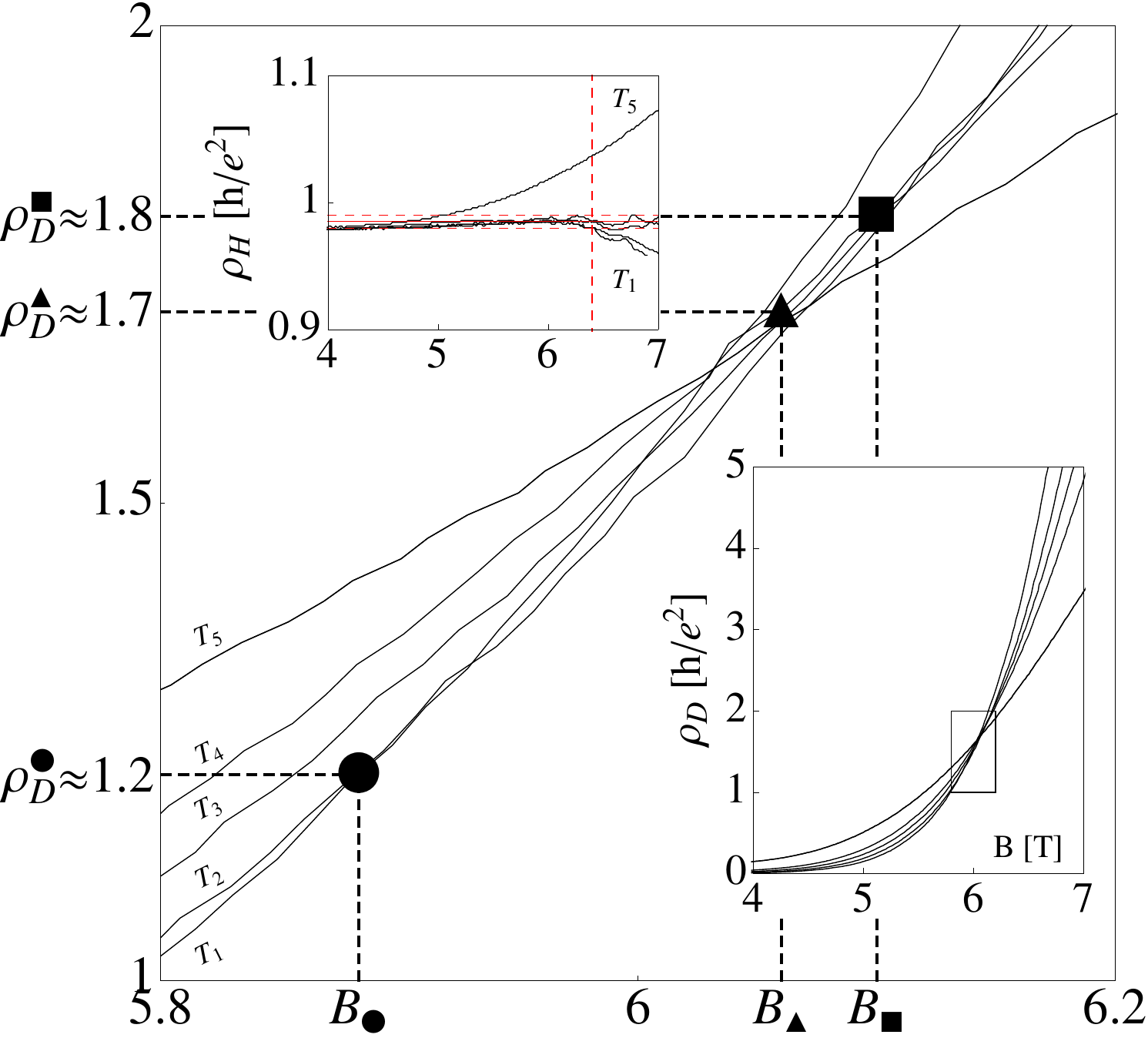}}
\end{center}
\caption[buprho]{Resistivity data\,\cite{Tsui_9810217,Tsui_9906212} for the $\nu = 1$ integer to insulator transition,
obtained for five different temperatures. Bottom inset:  the five experimental response curves  $\rho_D^{\phantom D}(B,T)$, with the 
region blown up in the main diagram outlined by a rectangle. 
Top inset: the five experimental response curves  $\rho_H^{\phantom D}(B,T)$ obtained in this experiment.}
\label{fig:Probes2-Fig3}
\end{figure}

\begin{figure}[t]
\begin{center}
{\includegraphics[scale = .5]{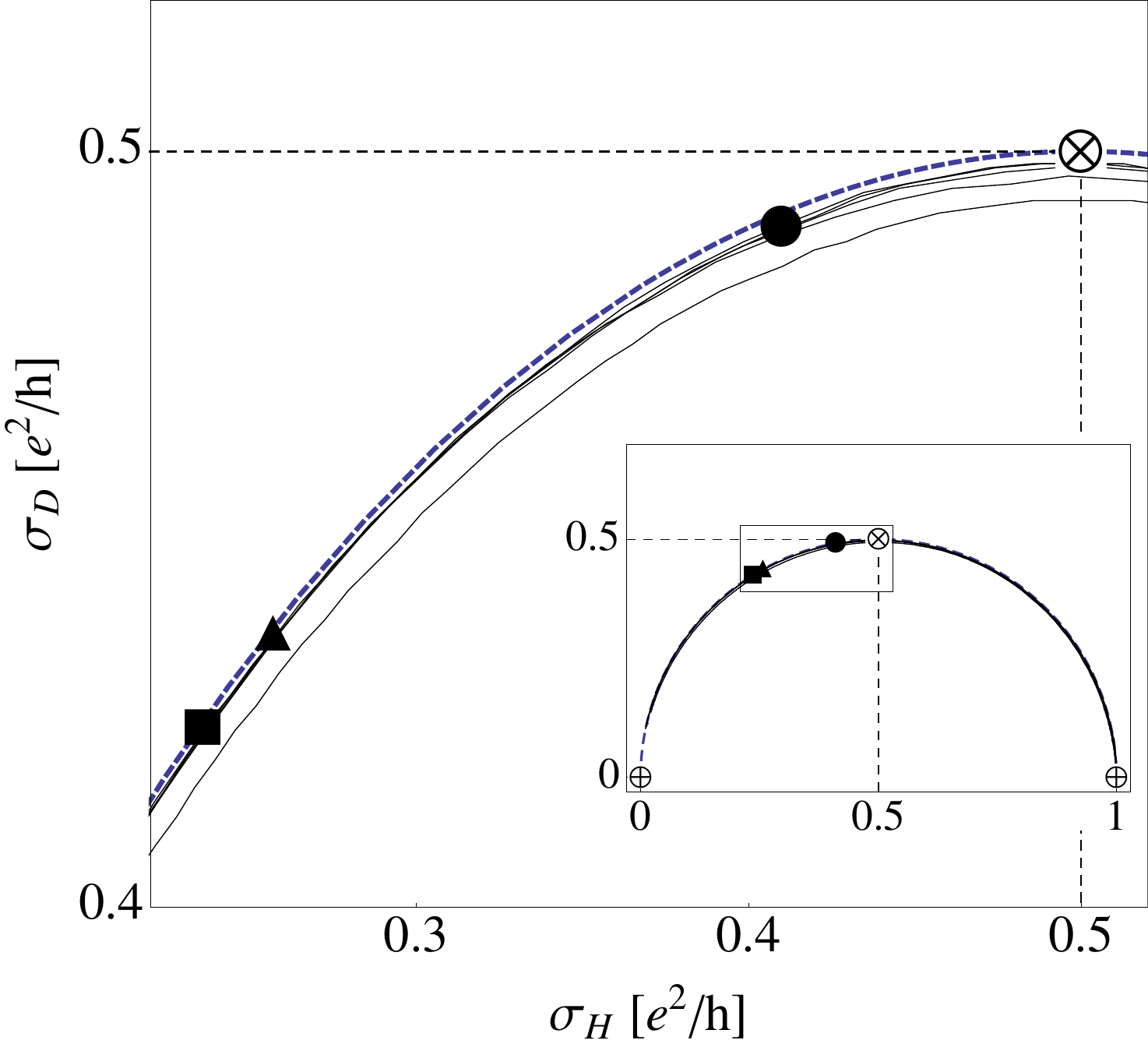}}
\end{center}
\caption[bupsigma]{Comparison of the complex conductivity $\sigma = \sigma_H^{\phantom D} + i \sigma_D^{\phantom D}$ 
calculated from the ``raw" resistivity data shown in 
Fig.\,\ref{fig:Probes2-Fig3}\,\cite{Tsui_9810217} with our holomorphic symmetry\,\cite{Oxford1}.
The region blown up in this diagram is indicated by a rectangle in the inset, and the three crossing points 
$(\bb, \blacktriangle, \bullet)$ from Fig.\,\ref{fig:Probes2-Fig3} are represented by the same icons.}
\label{fig:Probes2-Fig4}
\end{figure}

There have been attempts to explain this discrepancy, based on possible uncertainties in the
measurement method, or on nonlinear corrections\,\cite{Dolan}.  
However, a close examination of the data suggest a different and much simpler explanation.
An emergent quantum symmetry like $\Gamma_{\rm H}$ is always approximate, and can 
not be expected to give an accurate account of the physics unless all thermal fluctuations have been
effectively eliminated.  Comparing with the temperature range needed to identify the RG flow shown in Figs.\,\ref{fig:Probes2-Fig1} 
and \ref{fig:Probes2-Fig2}\,\cite{Oxford2} we expect it necessary to require temperatures below a few hundred milli-Kelvin. 
The implication of this is that the reported discrepancies with the critical conductivities is simply due 
to the fact that there is no common crossing point yet, 
because the samples were not cold enough to properly access the  quantum scaling regime. 

To justify this assertion we reproduce in Fig.\,\ref{fig:Probes2-Fig3}  the ``raw" resistivity data\,\cite{Tsui_9810217,Tsui_9906212} 
for the $\nu = 1$ integer to insulator transition,
obtained for temperatures  in the range $0.5-3.2 \,K$\,\cite{Tsui_9810217}\footnote{The temperature range $0.6 - 4.2\,K$ 
quoted in ref.\,\cite{Tsui_9906212} is presumably a misprint, as the data are identical. 
The precise values are not relevant to our discussion here, but the ordering is:  $T_1<T_2<T_3<T_4<T_5$.}. 
The bottom inset shows the five experimental (dissipative) response curves  $\rho_D^{\phantom D}(B,T)$, with the 
region blown up in the main diagram enclosed by a rectangle.

At first sight there appears to be a common crossing point of these five curves (lower inset), but closer examination (main diagram)
reveals that this is an ``optical illusion", and  that the crossing point is extremely sensitive to temperature.

Their first estimate\,\cite{Tsui_9810217}  of the critical field strength $B_{\blacktriangle} \approx 6.06\,T$ gives 
$\rho_{D}^{\blacktriangle} \approx 1.7 \, [h/e^2]$.
Their second estimate\,\cite{Tsui_9906212} $B_{\bb} \approx 6.1\,T$ gives $\rho_{D}^{\bb}\approx 1.8 \, [h/e^2]$.
This is 50\% higher than the value  $\rho_{D}^{\bullet} = \rho_D^{\phantom D}(B_{\bullet} \approx 5.88\,T ) \approx 1.2 \, [h/e^2]$ 
obtained from the 
intersection of the two lowest temperature curves, the latter being much closer to our theoretical self-dual/fixed point value 
$\rho_D^\otimes =  \rho_D^{\phantom D}(B_\otimes) = 1\, [h/e^2]$\,\cite{Oxford1}.  

There is no indication in the data shown in Fig.\,\ref{fig:Probes2-Fig3}  that the temperature 
is sufficiently low for $B_{\bullet}$ to be a good estimate of the true critical field strength $B_\otimes$, so we may
expect the crossing point to keep moving at still lower temperatures not accessed in this experiment. 
Given the significant measurement errors evident from the wiggles in the resistitivity traces, it is reasonable to 
conclude that  the data is not inconsistent with the prediction of the modular symmetry.

The Hall response seems to be less sensitive to the temperature.
The top inset in Fig.\,\ref{fig:Probes2-Fig3} shows the five experimental Hall curves  $\rho_H^{\phantom D}(B,T)$ obtained in this experiment.
For small $T\lesssim 1\,K$ and $B\lesssim 6.4\,T$ they are approximately constant, $\rho_H^{\phantom D} = 0.985\pm 0.005\, [h/e^2]$,
and very close to the theoretical value $\rho_H^\otimes =  \rho_H^{\phantom\otimes}(B_\otimes)  = 1\, [h/e^2]$ 
predicted by the holomorphic modular symmetry  $\Gamma^{\rm H}$\,\cite{Oxford1}.

For a more efficient ``global" comparison with our symmetry we exhibit in Fig.\,\ref{fig:Probes2-Fig4} 
the complexified conductivity\,\cite{Tsui_9810217}.
This is mathematically equivalent to the information contained in  Fig.\,\ref{fig:Probes2-Fig3}, but it provides a more intuitive
representation of the data, and one that starts to reveal the (hyperbolic) geometry built into our model.  

The actual data plotted in Fig.\,\ref{fig:Probes2-Fig4} is so similar to our modular phase diagram, 
represented by a dashed curve, that the two are only
distinguishable after a tenfold magnification of the small region enclosed by the rectangle in the inset.
At low temperatures the data collapse to a semi-circle, as predicted by our model\,\cite{Oxford1},
in which the semi-circle is the so-called ``hyperbolic geodesic" connecting the real fixed points.

Furthermore, the low temperature crossing point ($\bullet$) appears to be rapidly approaching
the  location of the quantum critical point $\sigma^{\phantom D}_\otimes = (1 + i)/2$, predicted by 
our model to be one of the elliptic fixed points of the holomorphic modular symmetry group $\Gamma_{\rm H}$\,\cite{Oxford1}.

If this temperature sensitivity is indeed the explanation of the disagreement between theory and experiment,  
experiments done at still lower temperatures should eventually see a stationary, common intersection point of 
the curves obtained at different temperatures, at a value corresponding to the theoretical fixed point 
$\rho_{\otimes}^{\phantom D} = 1 + i$.

\subsection{B.  Plateau-plateau quantum critical point}

An experiment of this type has been carried out recently\,\cite{Tsui_09050885}, 
in which the quantum phase transition
between the integer phases with fillings $\nu = 3$ and $\nu = 4$ was studied.
With temperatures ranging from  $510\,mK$ all the way down to $13\,mK$, 
this experiment has opened access to a new temperature regime
an order of magnitude lower than previous QH experiments.
This means physics can now be explored much deeper into the quantum domain,
where we expect $\Gamma_{\rm H}$ to reign.  
Experiments probing this domain will therefore subject the predictions of $\Gamma^{\rm H}$ to much more stringent tests than previously possible, 
and the experiment reported in ref.\,\cite{Tsui_09050885} is no exception.
Fortunately, it does in fact provide the most compelling evidence for our symmetry to date, with 
experiment and theory agreeing at the parts \emph{per mille} level.

\begin{figure}[t]
\begin{center}
{\includegraphics[scale = .9]{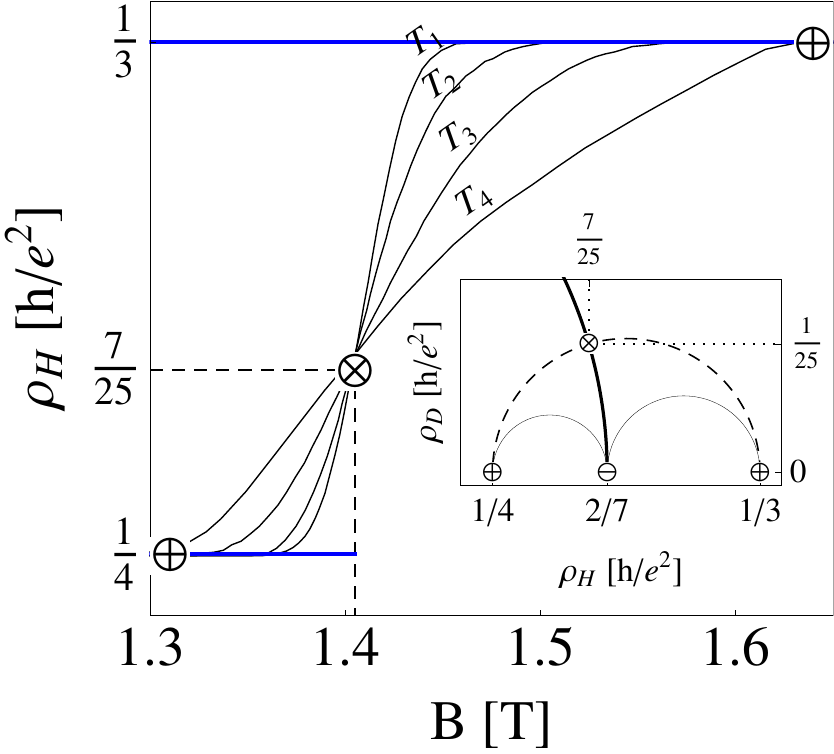}}
\end{center}
\caption[TsuiFig1]{Experimental traces of $\rho_H^{\phantom D}(B,T)$ 
for the quantum phase transition between the third and fourth integer plateaux
(adapted from ref.\,\cite{Tsui_09050885}), taken at the temperatures: $T_1 = 13,\,T_2 = 31,\,
T_3 = 114,\,T_4 = 510\,[mK] $.  
The graphs appear to cross in a single point with $B_\times = 1.405\pm 0.005\,T$, 
indicating that the scaling region has been reached.
Inset: resistivity phase diagram.}
\label{fig:Probes2-Fig5}
\end{figure}

The data are reproduced in Fig.\,\ref{fig:Probes2-Fig5}, but since the raw data have not been published 
the accuracy of our reproduction is limited by the pixel size used in Fig.\,1 of ref.\,\cite{Tsui_09050885}.
However, since the thickness of the published traces presumably has been chosen to obviate the need for error bars,
it is unlikely that we have introduced significant additional error by this semi-manual rendering.
Fig.\,\ref{fig:Probes2-Fig5} should therefore be a faithful representation of the real experimental data.

Fig.\,\ref{fig:Probes2-Fig5} shows four resistivity traces obtained at different temperatures 
which, unlike the case depicted in the previous figure, appear to be crossing in a single
point with $B_\times\approx 1.4\,T$.  This gives a  well defined critical value of the Hall resistivity 
that we estimate to be at $\rho^{\rm exp}_H (B_\times)= 0.282\pm 0.002\,[h/e^2]$.

The theoretical fixed point value following from $\Gamma^{\rm H}$ symmetry is in this case 
$\rho_H^{\otimes} =  7/25 = 0.28\,[h/e^2]$, which only differs from  $\rho^{\rm exp}_H$ 
by a few \emph{per mille}.  The two points are all but indistinguishable 
when plotted on the scale used in Fig.\,\ref{fig:Probes2-Fig5}.  

Observe that the only trace that appears slightly displaced from $\rho_\otimes$ 
is obtained at the highest temperature ($T_4 \approx 0.5\,K$), and that it exhibits a very poorly articulated Hall quantization.  
This suggests that the system is not deeply into the quantum domain when the temperature exceeds a few hundred milli-Kelvin,
and therefore is in a regime where the modular symmetry is expected to fail.
It is therefore rather surprising how accurate it still appears to be.  
Given a proper understanding of the separation of scales in this system,
it would be possible to estimate how accurate the effective field theory is, 
and therefore to quantify how approximate the symmetry is as a function of scale.
However, we are not yet able to do so.

The inset in Fig.\,\ref{fig:Probes2-Fig5} is a magnification of the relevant region of the modular phase diagram,
with a few simple (semi-circular) RG flow lines displayed. 
The dashed curve is a separatrix for the flow, and the heavy black curve is a 
phase boundary.   We see that the theory also predicts that this experiment should find 
$\rho_D^{\rm exp}(B_\times)\approx \rho_D^{\otimes} = 1/25 = 0.04\left[ h/e^2\right]$, 
but we have no published data to compare with this prediction.

\subsection{C.  RG flow}

In the neighbourhood of delocalization fixed points, \emph{i.e.}, near the quantum critical points 
labelled by $\otimes$ in our diagrams, the flow is found in numerical studies to be hyperbolic 
(anti-holomorphic)\,\cite{Oxford2}. This relies on the observation that, except for the sign, 
numerical simulations of the delocalization transition appear to give the same value for
the relevant and the irrelevant exponents. 

The relevant exponent obtained in these ``numerical experiments" agrees very well with the exponent
measured in real experiments\,\cite{nu-exp1,nu-exp2,nu-exp3,num1,num2}. 
The ``super-universal exponent" found for all phase transitions, both between integer and fractional levels,
was one of our original motivations for introducing $\Gamma_{\rm H}$.

Experimental information on the irrelevant exponent is contained in the temperature driven flows shown in 
Figs.\,\ref{fig:Probes2-Fig1} and \ref{fig:Probes2-Fig2}\,\cite{Oxford2}. 
The agreement is reasonable given the large experimental errors associated 
with the temperature measurements\,\cite{Murzin}. 

On the theoretical side the modular symmetry predicts that there should be super-universality of 
the scaling exponents associated with the various delocalization fixed points. 
It also predicts the anti-holomorphic scaling form, and indeed the predicted 
value of the relevant exponent following from $\Gamma_{\rm H}$ is in excellent agreement 
with the measured value\,\cite{Oxford2}. 

This pleasing agreement between theory and experiment has been challenged recently by an experiment probing the plateau-insulator  transition\,\cite{pruisken}. By reversing the magnetic field an irrelevant exponent was measured and found to be of magnitude significantly different from the relevant exponent. 
However, the temperatures used in this experiment are quite high, from $4.2\,K$ to $0.37\,K$. 
Indeed, as may be seen from Fig.\,2 in ref.\,\cite{pruisken}, their exponent is extracted from 
data taken at temperatures above $T \approx 1\,K$, the data below this being consistent with a zero exponent. 
This is significantly warmer than the range needed to find the quantum critical point (see Fig.\,\ref{fig:Probes2-Fig5}) 
in previous experiments and  raises the question whether this exponent is relevant to the quantum phase transition. 

\begin{figure}[t]
\begin{center}
\includegraphics[scale = .45]{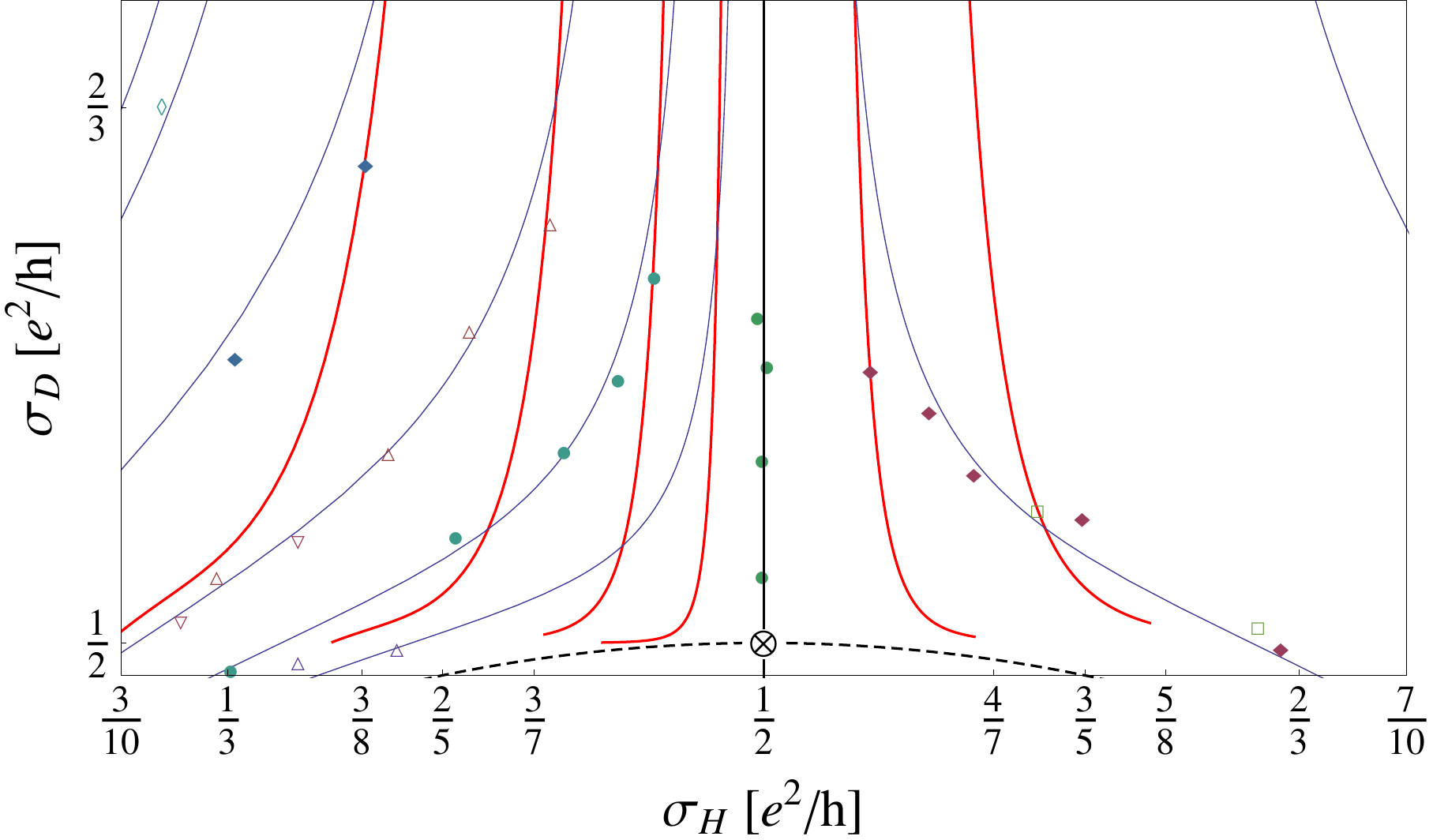}
\end{center}
\caption[endoflaw]{(Color online) Comparison of RG flows in the neighbourhood of the $\sigma_\otimes = (1/2,1/2)$ delocalisation fixed point.
The fat red flow lines, generated by the critical exponents found in ref.\,\cite{pruisken}, and the thin blue flow lines, 
generated using the critical exponents following from $\Gamma^{\rm H}$, should be compared with the experimental 
temperature driven data from ref.\,\cite{Grenoble1, Grenoble2} (discrete markers).}
\label{fig:Probes2-Fig6}
\end{figure} 

To test this we have plotted the RG flows near the $\sigma_\otimes^\ud = (1/2,1/2)$ saddle point shown in Fig.\,\ref{fig:Probes2-Fig6},
and compared these flows with the temperature driven flow data obtained in the Grenoble experiments\,\cite{Grenoble1, Grenoble2}.
We must choose one point on each flow line (initial value data needed to select a flow line), and we have
picked flow lines that pass near the first point of each sequence of data points (discrete markers) found  
experimentally\,\cite{Grenoble1, Grenoble2}.
The very asymmetric exponents from ref.\,\cite{pruisken} give the thick red flow lines.
Unlike the flow generated by using our exponents (thin blue lines), 
this flow is quite different from that found in the experiments, which were
performed at temperatures an order of magnitude lower than the experiments 
used to extract the asymmetric exponents,  with $T$ in the range from $0.3\,K$ down to $0.035\,K$.
This provides strong evidence that the experiment of ref.\,\cite{pruisken} is not probing the quantum phase transition regime,
and that it is therefore not a relevant experimental test of the modular symmetry predictions.

In summary, we find that the experimental data on the location of the delocalisation fixed points, 
as well as the geometry and absolute rate of RG flow near these points,  strongly supports 
the proposal of  ref.\,\cite{Oxford1},  that the QH system is symmetric under the holomorphic subgroup $\Gamma_0(2)$ 
of the emergent symmetry group $\Gamma_{\rm H}$.  
We also expect the QH system  to be invariant under the non-holomorpic symmetries generated by $J$, 
which taken together with the holomorphic symmetries leads to a web of dualities that we now discuss.

\section{IV.  PLATEAU-INSULATOR  DUALITY}

We consider first the transition between the QH insulator (QHI) phase, whose attractor in the conductivity plane 
is located at $\sigma_\oplus = 0$ (vanishing filling factor), 
and the neighboring integer phase corresponding to the plateau at $\sigma_{\oplus^\p} = 1$ 
(filling factor $\nu=1$), {\it c.f.} Fig.\,\ref{fig:Probes2-Fig1}.  
Attractive fixed points (the plateaux) are all located on the real line (``compactified"
by adding the ``point" $i\infty$) and labelled by the icon $\oplus$ in our diagrams. 
The repulsive fixed points are similarly labelled by the icon $\ominus$ in our diagrams. 
We use the notation $0\otimes_{\sigma}1$ to denote this very symmetric transition. 
The transition between phases is driven by a change in the applied magnetic field $B$.   
The dynamics associated with the holomorphic symmetry $\Gamma_0(2)$ forces the system along the semi-circle connecting 
the plateaux\,\cite{Oxford1, Cliff}, through the quantum critical point at 
$\sigma_\otimes = (1 + i)/2$.

The corresponding resistivity transition connects the attractors at  $S(1)  = -1$ and $S(0) = i\infty - 1$,
i.e., the plateaux in the Hall resistivity at $1$ and $\infty$, via a quantum critical point at  $\rho_\otimes = i - 1$.

The circular symmetry of the $0\otimes_{\sigma}1$ transition suggests that an angular parameterization may be useful, so we set $\sigma(\theta) \stackrel{\cap}{=} (1 + \exp(i\theta))/2$ with $\theta\in[0,\pi]$.
The equal sign capped by a semi-circle ($\stackrel{\cap}{=}$) is a pictogram used to remind us that the 
expression is only to be used along the semi-circle connecting $\oplus$ and 
$\oplus^{\p}$.   The fixed points appear at the angles $\theta = 0, \pi/2, \pi$:  $\sigma(0) = 1 = \oplus^{\p},\;
\sigma(\pi /2) =  (1+i)/2 = \otimes$ and $\sigma(\pi) = 0 = \oplus$.  

Note that in this particular case the simple symmetry of the semi-circle is enough to infer that the duality transformation 
preserves the value of $\sigma_D^{\phantom D}$ and reflects the value of $\sigma_H^{\phantom D}$ in the 
vertical symmetry axis $\sigma_H^{\phantom D} = 1/2$.  
In other words, $\sigma^d(\theta) \stackrel{\cap}{=} \sigma(\pi - \theta)$.

All other transitions are images of this one under some transformation in the symmetry group 
$\Gamma_{\rm H}$.  The easiest way to find this transformation is to exploit one of the most surprising
facts about modular groups:  starting anywhere in the upper half of the complex plane \emph{the flow can 
only end at a rational number} on the real line, provided it avoids hitting one of the semi-stable quantum critical points
where it gets stuck.   This was originally the most compelling reason for considering modular groups\,\cite{Oxford1}, 
since it immediately encodes the extremely accurate quantization of the Hall resistivity 
at rational values as an automatic consequence of (an emergent) symmetry.

Consider therefore how modular transformations act on the rational numbers.  
If we represent a fraction $f = p/q$ by the column vector\footnote{In order for our formulas 
to be valid for all fixed points we adopt the convention that $0 = (0,1)$ and $i\infty = (1,0)$.} 
$F = (p,q)^t$, and a modular transformation 
$\gamma\in\Gamma_0(2)$ by the matrix in eq.\,(\ref{eq:modulartrafo}), then the image of 
$f$ under $\gamma$ is given by the matrix product $\gamma\cdot F$.
A direct transition  between the phases labelled by $\oplus = f$ and $\oplus^{\p} = f^{\p}$ is possible only 
if there is a modular transformation mapping the semi-circle $0\otimes_{\sigma}1$ to the semi-circle 
$f\otimes_{\sigma} f^{\p}$.  This gives $\gamma = (F^{\p} - F, F)$, whence the transition is allowed iff 
\begin{equation}
\det\gamma = \det(F^{\p},F) = p^{\p} q - p q^{\p} = 1\;.
\label{eq:det}
\end{equation}

Using $\gamma$ we find that the angular parametrization of the $0\otimes_{\sigma}1$ transition maps to:
\begin{equation*}
\sigma_{\rm ff^\p}^{\ud}(\theta) \stackrel{\cap}{=}
\frac{p\sin(\theta/2) + i p^\prime\cos(\theta/2)}{q\sin(\theta/2) + i q^\prime\cos(\theta/2)}\;.
\end{equation*}
Again we find the attractive fixed points at $\theta = 0, \pi$:
\begin{equation*}
\sigma_{\rm ff^\p}^{\ud}(0) = f^{\p} = \oplus^{\p}\;,\;\sigma_{\rm ff^\p}^{\ud}(\pi) = f = \oplus\;,
\end{equation*}
and the saddle point at $\theta = \pi/2$:
\begin{equation*}
\sigma_{\rm ff^\p}^{\ud}(\pi /2) = \frac{p + i p^{\p}}{q + i q^{\p}} = \otimes\;.
\end{equation*}
The duality transformation is still given by the simple expression: 
\begin{equation}
\sigma_{\rm ff^\p}^d(\theta) \stackrel{\cap}{=} \sigma_{\rm ff^\p}^{\ud}(\pi - \theta)\;.
\label{eq:semidual}
\end{equation}
Decompressing this constrained duality transformation into components we obtain:
\begin{eqnarray}
\sigma_H^d &\stackrel{\cap}{=}&
\frac{(p^{\p 2} q^{\p 2} - p^2 q^2) + (pq^3 - p^\p q^{\p 3})\sigma_H^{\phantom D} }
{(p^\p q^{\p 3} -  p q^3) + (q^4 - q^{\p 4})\sigma_H^{\phantom D}}
\label{eq:ourLaw}\\
\sigma_D^d &\stackrel{\cap}{=}& \frac{qq^\p \sigma_D^{\phantom D}}
{(p^\p q^{\p 3} - p q^3) + (q^4 - q^{\p 4}) \sigma_H^{\phantom D}}\;\;.
\end{eqnarray}
Notice that while $\sigma_D^d$ in general depends on both 
$\sigma_H^{\phantom D}$ and $\sigma_D^{\phantom D}$ (except when $q^\prime = q$, or $q^\prime = 0$), 
$\sigma_H^d$ always decouples from $\sigma_D^{\phantom D}$ along the semi-circle.

It is not clear to us whether experimental tests of such generalized dualities, 
connecting two fractional phases, say, are feasible.  
Here we focus on the small subset of these transformations (still infinite) involving 
any transition to the Hall insulator phase, labelled in the $\sigma$-plane by $(p,q) = (0,1)$.
A direct transition from the insulator phase to another phase is only possible if the modularity condition
eq.\,(\ref{eq:det}) is satisfied, from which we derive that the neighboring phases in the $\sigma$-plane
are labelled by $(p^\p, q^\p) = (1,m)$, with $m$ restricted to any odd integer since we are considering 
the sub-group $\Gamma_0(2)$ rather than the full modular group.  

Now the duality relations collapse and are best given in terms of the resistivity:
\begin{equation}
\rho_H^d = \rho_H^{\phantom D} = m\;,\quad \rho_D^d = 1/\rho_D^{\phantom D} \;.
\label{eq:rhoduality}
\end{equation}
We see that the characteristic features of modular duality involving the insulator phase, 
are that the diagonal resistivities are inversely proportional,
while the Hall resistance is independent of both $B$ and $T$, 
taking the same constant (odd integer) value $m \,[h/e^2]$ in both phases.  

\begin{figure}[t]
\begin{center}
{\includegraphics[scale = 0.8]{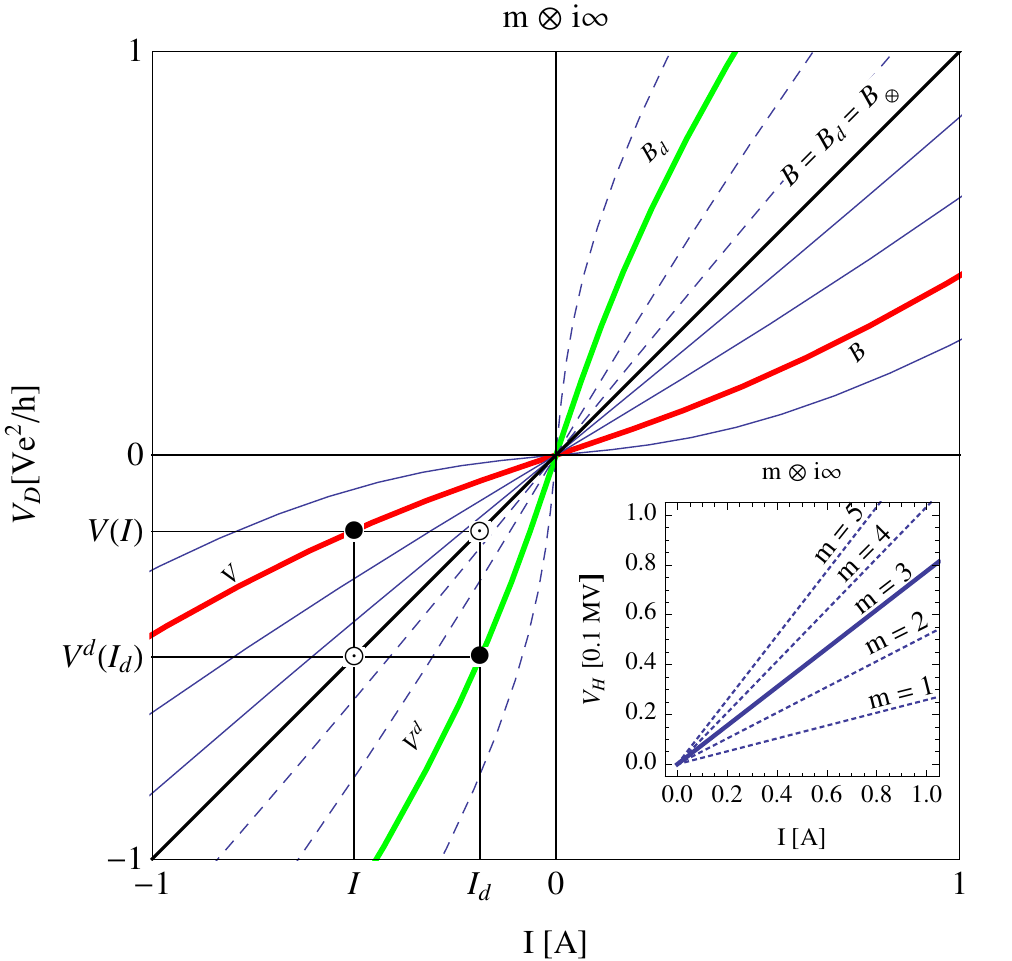}}
\end{center}
\caption[IV1]{(Color online) IV-characteristics for plateaux-insulator transitions $m\otimes_{\rho} i\infty$ ($m$ odd).
Main diagram:  schematic of $V_D(B,I) = R_D(B,I)\cdot I$ in units $[V e^2/h\sim A]$ where $R_D$
is dimensionless, for various dual pairs $(B,B_d)$ of magnetic fields.  The graphic in the third quadrant 
shows geometrically why $V$ and $V^d$ are inverse functions.
Bottom inset:  Predicted  $V_H$-traces according to the $\Gamma^{\rm H}$ duality relation for various plateau-insulator transitions.}
\label{fig:Probes2-Fig7}
\end{figure}

It is possible to test these duality relations, as far from the quantum critical point as one wants,
by choosing various magnetic field values in such a way that the
electro-magnetic response of the system in both phases can be recorded by taking IV-traces.
One then looks for magnetic field values $B_d$ dual to $B$
such that the IV-characteristics collapse to the same function of the current after 
transforming $\rho$ to its dual value $\rho^d$.  This is illustrated schematically in 
Fig.\,\ref{fig:Probes2-Fig7}, which for the case $m=3$ is to be compared with the real Hall data 
obtained in the experiment discussed next.
 
\section{V.  EXPERIMENTAL TEST OF DUALITY}

\begin{figure}[t]
\begin{center}
\includegraphics[scale = .8]{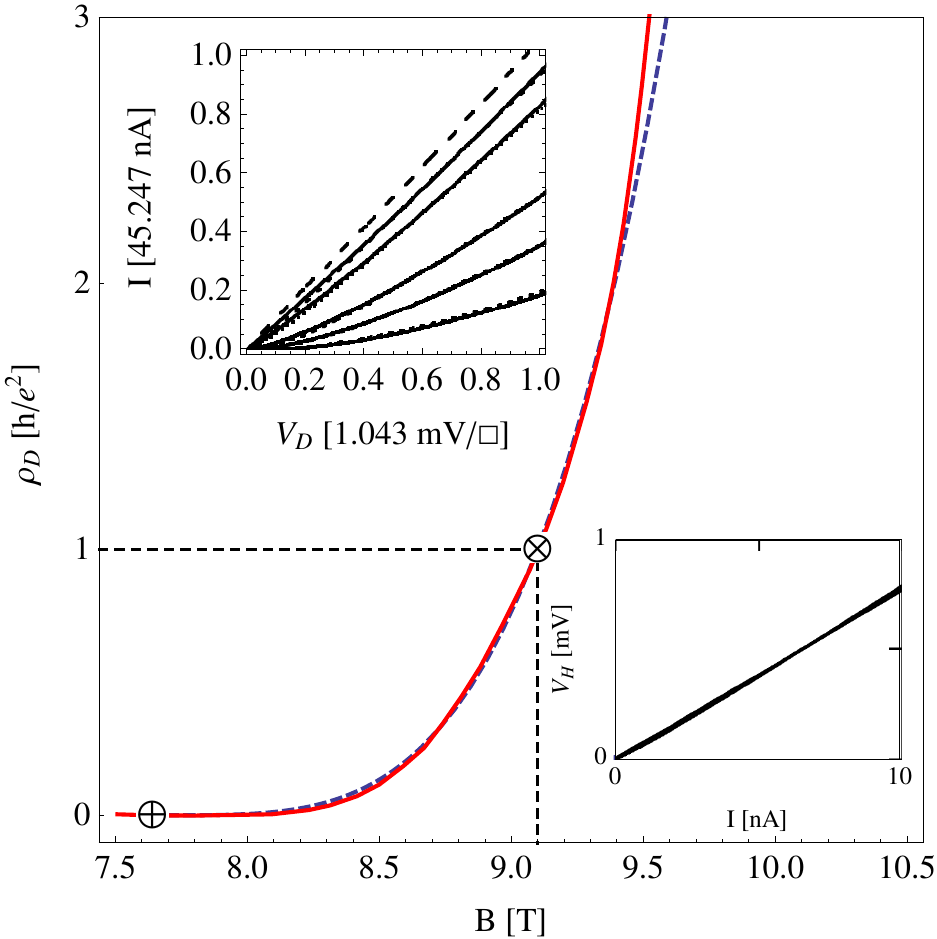}
\end{center}
\caption[fittedrho]{(Color online) Experimental discovery of duality in the QH system (adapted from ref.\,\cite{Tsui1}).
The solid red curve is a reproduction of the lowest temperature resistivity curve $\rho_D^\ud(B)$,obtained at $T = 26 \,mK$.
The dashed black curve shows $\rho_D^{\rm fit} (B) = (\Delta B/\Delta B_\otimes)^a$, fitted to the data with $a = 3.8$.
\emph{Top inset:} IV-traces recorded on both sides of the liquid-insulator quantum phase transition, obtained at  $T = 21\,mK$ 
and dual $B$-field values in  the range $8.5 - 10.1\,T$.
The low field data $(\nu = 1/3)$ have been reflected in the symmetry axis $V_D = I$
and are practically indistinguishable from the high field data,
showing that the five dual pairs of IV-characteristics are mirror images (inverse functions).
\emph{Bottom inset:}  The eleven $\rho_H^\ud(I)$-traces recorded in this experiment, 
with $B$-field values incremented in steps of $0.1\,T$ from $8.7\,T$ to $9.7\,T$, 
are indistinguishable and in agreement with the modular expectation when $m = 3$ (inset in Fig.\,\ref{fig:Probes2-Fig7}).}
\label{fig:Probes2-Fig8}
\end{figure}

In a remarkable experiment\footnote{The Hall bar is a high mobility $(\mu = 5.5 \times 10^5\,cm^2/sV)$, 
low density $(n = 6.5 \times 10^{10} \,/cm^2)$  GaAs/AlGaAs hetero-structure, with 
$R_D^c \approx 23 \;k\Omega$ and $\rho_{xx}^c  \approx 1\;[h/e^2]$\,\cite{Tsui2}.} 
Shahar et al.\,\cite{Tsui1} found clear evidence for a duality between 
current-voltage (IV) characteristics obtained on opposite sides of the QH liquid to insulator transition.  
They identified six pairs $(B, B_d)$ of dual magnetic field values, 
with $B$ and $B_d$ on opposite sides of the quantum phase transition at\footnote{Since no errors are given in ref.\,\cite{Tsui1}, 
our least biased estimate is to take the largest error consistent with the published value: $B_c = 9.1\pm 0.05\,[T]$,
and similarly for the other values of $B$.}  $B = B_c\approx 9.1\,T$,
separating the $\nu = 1/3$ fractional QH liquid from the QH insulator phase ($\nu = 0$).
With $B$-values in this range the transverse IV-characteristics $V_y(B,I) = R_{yx}(B,I) \cdot I$ 
was found to be independent of $B$ and linear in $I$, with slope  $R_H = R_{yx} \approx 3\;[h/e^2]$. 
We have reproduced their data showing this in Fig.\,\ref{fig:Probes2-Fig8}. 
The critical value of the Hall resistivity is therefore $\rho_H^c = \rho_{yx}(B_c) \approx 3\;[h/e^2]$.  

By contrast the dissipative IV-characteristics $V_x(B,I) = R_{xx}(B,I)\cdot I$ 
are extremely non-linear in both phases, degenerating to a linear (Ohmic) relation only when $B\rightarrow B_c$
(see insets in Fig.\,\ref{fig:Probes2-Fig8}). 

Shahar et al.\,\cite{Tsui1} discovered that to each $B$ there exists a dual field value $B_d$, such that the dissipative IV-curve 
$V(B,I)$ (suppressing the now superfluous subscript on $V_x$) after reflection in the diagonal $V = I$ is virtually identical to 
the dual IV-curve $ V_d(B_d,I_d)$ in the opposite phase.  Their diagrams proving this are reproduced in
Fig.\,\ref{fig:Probes2-Fig6}.  This means that $V$ and $V^d$ are inverse functions of $I$, which implies that
\begin{equation}
\rho_D^d(B_d,I_d) = 1/\rho_D^{\phantom D}(B,I)\;\; .
\label{eq:RRd}
\end{equation}

This duality structure is just that predicted by the modular symmetry,.
For the $\nu = 0$ to $\nu = 1/3$ transition $(p^\p, q^\p) = (1, 3)$,  and it follows from eq.\,({\ref{eq:rhoduality}) that 
$\rho_H^{d} = 3$, in agreement with the experimental result (bottom inset, Fig.\,\ref{fig:Probes2-Fig8}).
Clearly the modular duality relation for the dissipative part of eq.\,({\ref{eq:rhoduality}) is also in perfect agreement
with the experimental dissipative duality relation, eq.\,(\ref{eq:RRd}). 
In a transition to the insulator phase there must of course be pairs of points with inverse values of the resistivities,
but duality implies that the physics of the system at dual parameter values is the same, 
explaining why the full nonlinear structure of the IV-curves evident in Fig.\,\ref{fig:Probes2-Fig7} coincide so precisely.

As mentioned above the original interpretation\;\cite{Tsui1, nu_duality} of the duality observed in\,\cite{Tsui1} was that it followed from the law of corresponding states\,\cite{KLZ}. However this is not the case and in fact the experiment strongly disfavours the duality relation thus derived. To discuss this and to clarify the difference between the two forms of duality it is instructive to consider the structure in the conductivity plane.

For the $\nu = 0$ to $\nu = 1$ transition the corresponding states  are related by particle-hole duality giving $\nu_d = 1 - \nu$. 
The flux attachment transformation\;\cite{KLZ} $1/\nu^{\prime} = 1/\nu + 2 m$ ($m$ integer) maps this transition to the $\nu = 0$ to $\nu = 1/k$ 
transition, with $k = 2 m + 1$, giving the duality relation:  $1/\nu_d - k = (1/\nu - k)^{-1}$. 
For the $\nu = 0$ to $\nu = 1/3$ transition this gives\,\cite{Tsui1,nu_duality}  
\begin{equation}
\nu_{d}(\nu)  = \frac{1 - 3\nu}{3 - c_\nu \nu}\quad (c_\nu = 8)\;,
\label{eq:vvd}
\end{equation}
which relates the filling $\nu = \nu(B)$ to a dual filling $\nu_d = \nu(B_d)$.

This structure should be compared with the duality obtained from the modular symmetry given by  eq.\,(\ref{eq:ourLaw}). For the transverse conductivity we have
\begin{equation}
\sigma^d_H  = \frac{1 - 3 \sigma_H^\ud}{3 - c_m \sigma_H^\ud}\quad (c_m = 80/9)\;.
\label{eq:ssdxy}
\end{equation}

This is superficially similar to eq.\,(\ref{eq:vvd}), but these two transformations are in fact very different.
Not only does modularity ($\det \sigma_d = 1$) fix $c = 10$, but more importantly the underlying complex duality relation  
eq.\,(\ref{eq:semidual}) is much stronger and leads to the set of duality relations given in eq.\,(\ref{eq:rhoduality}). 

By comparison, the duality derived from the law of corresponding states has nothing to say about  the Hall response of the system, and does not predict the fact that it remains constant at its critical value 
$\rho_H^c \approx 3\,[h/e^2]$ across the transition. 
It is also not clear that eq.\,(\ref{eq:RRd}) relating dual values of $\rho_D^\ud$ applies for the the dual pairs of 
filling factors given by eq.\,(\ref{eq:vvd}), because the mapping used to derive the latter strictly applies only on 
the plateaux where $\rho_D^\ud$ vanishes. Moreover the self-dual point of this transformation is $\nu_* = 0.25$, 
distinct from the critical value  $\nu_c \approx 0.28$ found 
experimentally\,\cite{Tsui2}. In fact this form of duality does {\it not} reproduce the measured values of the dual filling factors. 
This is most easily demonstrated  by comparing the observed insulator values of $\Delta\nu = \nu - \nu_c$ reported in ref.\,\cite{Tsui1},
with the images of the low-field data (obtained in the $\nu = 1/3$ phase) under the transformation following immediately 
from eq.\,(\ref{eq:vvd}):
\begin{equation}
\Delta\nu_{d}(\Delta\nu)  = - \frac{(1 - 6\nu_c + 8\nu_c^2) + (8\nu_c - 3)\Delta\nu}{(8\nu_c - 3) + 8\Delta\nu}\;.
\label{eq:dvdvd}
\end{equation}
The image points do not coincide\,\cite{Oxford3} with the insulator fillings  reported in ref.\,\cite{Tsui1}.

In summary, the experimentally observed duality is in clear disagreement with the form obtained using 
the law of corresponding states, and in excellent agreement with the structure predicted by the modular symmetry $\Gamma_{\rm H}$. 
In the next section we consider whether it is possible in the latter case to predict the dual magnetic field values as well as the dual 
resistivities or conductivities.

\section{VI.  DUAL MAGNETIC FIELD VALUES}

 The modular symmetry by itself does not predict the values of the dual magnetic fields, but the duality transformation
 $B_d(B)$ must inherit some properties from the modular duality transformation, eq.\,(\ref{eq:rhoduality}), 
in particular how it treats the RG fixed points.  We also expect that a suitable first order Pad\'e approximation, 
of the fractional linear form 
\begin{equation}
 B_{d}(B)=\frac{aB-b}{B-c}\;\;,
 \label{pade}
 \end{equation}
 with $a,b$ and $c$ constants, should give a good representation of the duality transformation.
 This form has a simple pole at $B=c$ corresponding to the need to have 
 dual field values deep in the insulator phase. It also has the property that its form is maintained when expressing 
 $B$ as a function of $B_{d}$ as is required from the duality property of the theory. We can fix the three constants 
 by requiring that the self dual point should be at $B=B_{\otimes}$, and by demanding that the 
duality relation swap the plateaux value, $B=B_{\oplus}\;(\nu=1/m)$, with the insulator value,  $B=\infty\;(\nu = 0)$.
The resulting duality relation  has the form 
\begin{equation}
B_d(B) =\frac{B B_\oplus  + (B_\otimes - 2B_\oplus) B_\otimes}{B - B_\oplus}\;\;.
\label{eq:BBd}
\end{equation}
Both the parameters $B_\otimes$ and $B_\oplus$ are of course non-universal, 
but can be extracted from each experiment as follows.
$B_\otimes \approx B_\times$ is the critical field value, identified in the experiment from the 
temperature independent crossing point of all the resisitivity traces.  
$B_\oplus$ is given by the measured value of  $B$ at the centre of the plateau.

We can write the $B$-duality relation in a universal form that can be easily tested when duality is 
investigated for  other plateau-insulator  transitions.
Start by introducing the ``reduced" $B$-field $b = (B - B_\otimes)/B_\otimes$, which is analogous to the reduced temperature 
$t = (T -T _c)/T_c$ used in the study of classical phase transitions.
We also choose to measure all fields in the unit $\alpha =  - b_\oplus = (B_\otimes - B_\oplus)/B_\otimes > 0$,
whence $b = \beta\alpha \equiv \beta$ and  $b_d = \beta_d\alpha \equiv \beta_d$.
With this notation all non-universal (system- and transition-specific) data have been absorbed in the units, 
and the $B$-duality transformation takes the ``super-universal" form: 
\begin{equation}
\beta_d(\beta) = - \frac{\beta}{1 + \beta}\;\;.
\label{universal}
\end{equation}
The prediction is now that the dual $B$-field data for any plateau-insulator transition will collapse onto this curve.

As discussed above there is a measurement\,\cite{Tsui1}  of the dual $B$ pairs for the transition from the $\nu=1/3\,( m=3)$ phase. 
This experiment found $B_\otimes = 9.1\,T$, and we estimate that $B_\oplus \approx 7.6\,T$,  corresponding to the centre of the $\nu=1/3$ plateau. (For $\nu\propto B^{-1}$ this value also agrees very well with the positions of the other plateaux.)
Using these values of the parameters we have transposed the pairs $(B,B_d)$ found in ref.\,\cite{Tsui1} to the $\beta$-basis and plotted 
the pairs $(\beta, \beta_d)$ as points with appropriate error bars in Fig.\,\ref{fig:Probes2-Fig9}.
The universal curve given by eq.\,(\ref{universal}) is shown as a solid black curve  in Fig.\,\ref{fig:Probes2-Fig9}, 
and we see that the handful of available experimental data points are in excellent agreement with the 
form derived from eq.\,(\ref{eq:BBd}).

One way to see why this approximation gives such a good result
is to observe that a simple power law gives an accurate model of the resistivity data.  The function
\begin{equation}
\rho_D^{\rm fit} (B) = (\Delta B/\Delta B_\otimes)^a \;\;,
\end{equation}
with $\Delta B = B - B_\oplus$, $\Delta B_\otimes = B_\otimes - B_\oplus$ and the fitted value $a\approx3.8$,
is shown as the dashed curve in Fig.\,\ref{fig:Probes2-Fig8}. We see that it gives a good fit to the real resistivity trace
(solid red curve, adapted from the lowest temperature trace in ref.\,\cite{Tsui1}) for a reasonable range of $B$-fields,
and only deviates when $B$ goes deeply into the insulator phase $(B \rightarrow \infty)$.
If we set $\rho_D^{\rm fit} (B_d) = 1/\rho_D^{\rm fit} (B)$, the fitting parameter drops out and eq.\,(\ref{eq:BBd}) follows.

Since duality relates all the plateau-insulator transitions, we expect the fractional linear approximation in  
eqs.\,(\ref{eq:BBd}) and (\ref{universal})
to work equally well for any transition of this type. It should therefore provide an accurate prediction for dual pairs of 
$B$-fields across any plateau-insulator  $(\nu = 1/m\leftrightarrow\nu = 0)$  quantum phase transition.

\begin{figure}[t]
\begin{center}
\includegraphics[scale = .45]{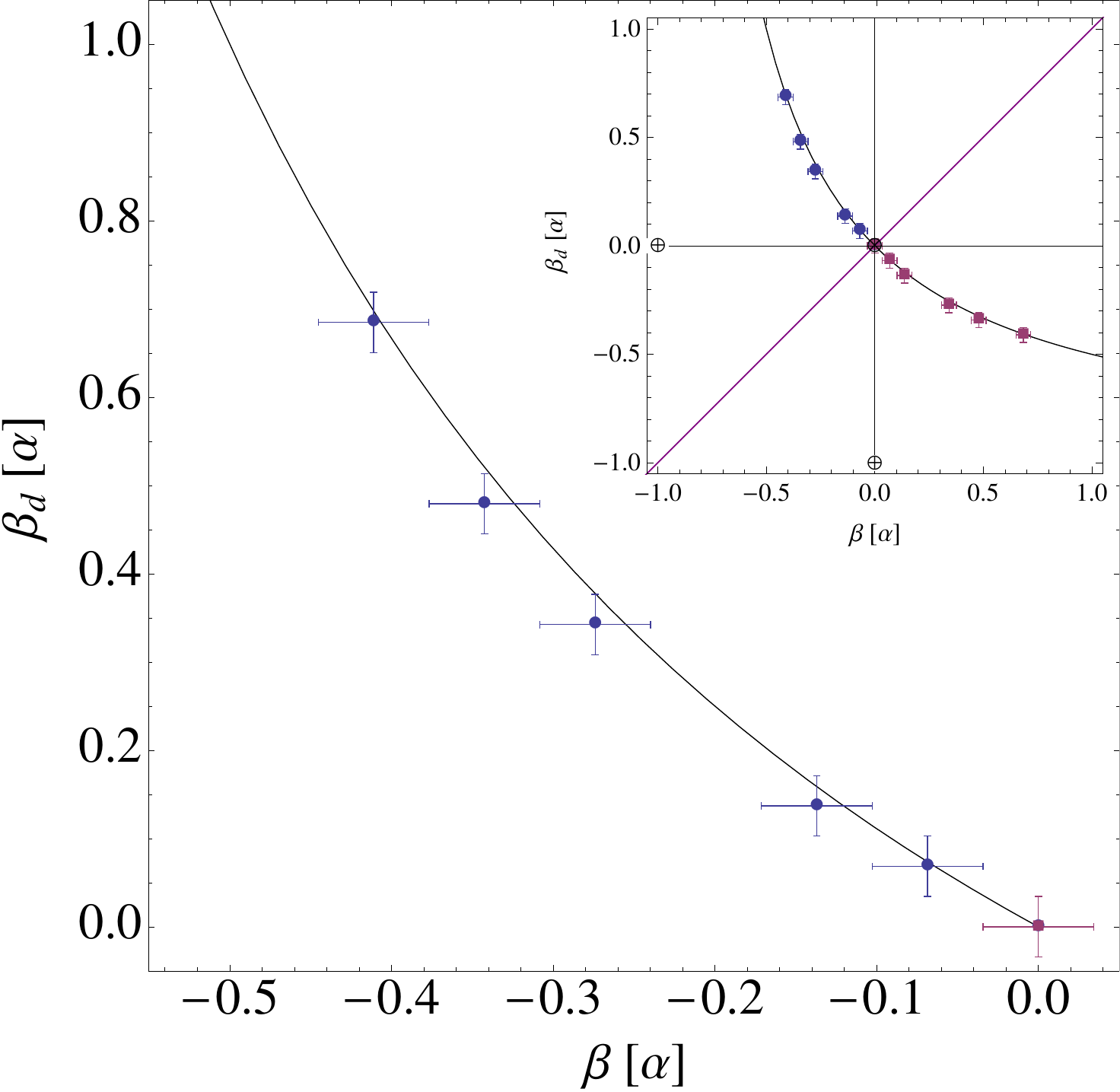}
\end{center}
\caption[endoflaw]{(Color online) Universal curve for $B$-field duality across any plateau-insulator transition, 
at leading order in the Pad\'e-expansion.  The only available duality data on $B$-fields is taken from ref.\,\cite{Tsui1}.}
\label{fig:Probes2-Fig9}
\end{figure} 

\section{VI.  SUMMARY}

The modular quantum Hall symmetry group\,\cite{Oxford1}  $\Gamma_{\rm H} = \mathrm{Aut}\,\Gamma_0(2)$ successfully 
predicts \emph{all} major features observed in fully spin-polarized quantum Hall experiments:\hfill\break
the integer and (odd-denominator) fractional quantization of the Hall resistance;\hfill\break
the exact location of quantum critical points in the complex conductivity (resistivity) plane, for both plateau-plateau and 
plateau-insulator transitions;\hfill\break
the geometry of RG flows;\hfill\break
the rate of RG flows, including the observed ``super-universality" of the critical exponents;\hfill\break
and finally, the non-linear dualities relating IV-characteristics measured in different phases, 
on opposite sides of quantum phase transitions.\hfill\break

Available experimental data on RG flows  are beginning to provide a detailed
phase portrait of the QH system, which is in impressive agreement with the anti-holomorphic flow 
derived from  $\Gamma_{\rm H}$ (Figs.\,\ref{fig:Probes2-Fig1} and \ref{fig:Probes2-Fig2}).

A close examination of magneto-transport data reveals a hyper-sensitivity to temperature that has not 
received sufficient attention (Figs.\,\ref{fig:Probes2-Fig3} and \ref{fig:Probes2-Fig4}), and these data therefore do \emph{not} invalidate the  
quantum symmetry $\Gamma_{\rm H}$.
On the contrary, at sufficiently low temperatures (below a few hundred $mK$ for the experiments reviewed here) 
$\Gamma_{\rm H}$ predicts the location of a quantum critical point that agrees with experiment at the parts \emph{per mille} level
(Fig.\,\ref{fig:Probes2-Fig5}).

The remarkable experimental duality relation between IV curves obtained for pairs of dual magnetic field values\,\cite{Tsui1},
 provides unambiguous evidence of a duality symmetry in the QH system, for a particularly simple 
 quantum phase transition (Fig.\,\ref{fig:Probes2-Fig8}).
 It reveals a profound connection between the transport mechanisms deep inside the 
 quantum liquid and insulator phases, a symmetry which holds to great accuracy even far from the quantum critical point at
$(\rho_H^c, \rho_D^c)  \approx  (3,1)\;\;[h/e^2]$.

There is a significant difference between the quality of the predictions for the dual pairs of filling factors derived from 
the so-called ``law of corresponding states", and the predictions for the dual resistivities and dual field values derived 
from the modular symmetry acting on the complexified resistivity.
While the former is strongly disfavored by experiment, the agreement between the modular predictions and the
data obtained so far are striking.

In anticipation of a new generation of milli-Kelvin QH experiments, pioneered in ref.\,\cite{Tsui_09050885}, we have 
worked out generalized modular duality transformations for the conductivities (resistivities) and magnetic fields.
These experiments will finally, two decades after its inception, subject our proposal for an emergent quantum Hall
symmetry $\Gamma_{\rm H}$ to rigorous tests.  Since this symmetry is completely rigid these
tests will be conclusive,  if they are carried out.


\end{document}